\documentclass[10pt,a4paper,onecolumn]{article}
\usepackage{marginnote}
\usepackage{graphicx}
\usepackage[rgb]{xcolor}
\usepackage{authblk,etoolbox}
\usepackage{titlesec}
\usepackage{calc}
\usepackage{tikz}
\usepackage[pdfa]{hyperref}
\usepackage{hyperxmp}
\hypersetup{%
    unicode=true,
    pdfapart=3,
    pdfaconformance=B,
    pdftitle={Eureka!: An End-to-End Pipeline for JWST Time-Series
Observations},
    pdfauthor={Taylor J. Bell, Eva-Maria Ahrer, Jonathan Brande, Aarynn
L. Carter, Adina D. Feinstein, Giannina Guzman Caloca, Megan Mansfield,
Sebastian Zieba, Caroline Piaulet, Björn Benneke, Joseph Filippazzo,
Erin M. May, Pierre-Alexis Roy, Laura Kreidberg, Kevin B. Stevenson},
    pdfpublication={Journal of Open Source Software},
    pdfpublisher={Open Journals},
    pdfissn={2475-9066},
    pdfpubtype={journal},
    pdfvolumenum={7},
    pdfissuenum={76},
    pdfdoi={10.21105/joss.04503},
    pdfcopyright={Copyright (c) 2022, Taylor J. Bell, Eva-Maria Ahrer,
Jonathan Brande, Aarynn L. Carter, Adina D. Feinstein, Giannina Guzman
Caloca, Megan Mansfield, Sebastian Zieba, Caroline Piaulet, Björn
Benneke, Joseph Filippazzo, Erin M. May, Pierre-Alexis Roy, Laura
Kreidberg, Kevin B. Stevenson},
    pdflicenseurl={http://creativecommons.org/licenses/by/4.0/},
    colorlinks=true,
    linkcolor=[rgb]{0.0, 0.5, 1.0},
    citecolor=Blue,
    urlcolor=[rgb]{0.0, 0.5, 1.0},
    breaklinks=true
}
\immediate\pdfobj stream attr{/N 3} file{sRGB.icc}
\pdfcatalog{%
  /OutputIntents [
    <<
      /Type /OutputIntent
      /S /GTS_PDFA1
      /DestOutputProfile \the\pdflastobj\space 0 R
      /OutputConditionIdentifier (sRGB)
      /Info (sRGB)
    >>
  ]
}
\usepackage{caption}
\usepackage{orcidlink}
\usepackage{tcolorbox}
\usepackage{amssymb,amsmath}
\usepackage{ifxetex,ifluatex}
\usepackage{seqsplit}
\usepackage{xstring}

\usepackage{float}
\let\origfigure\figure
\let\endorigfigure\endfigure
\renewenvironment{figure}[1][2] {
    \expandafter\origfigure\expandafter[H]
} {
    \endorigfigure
}

\let\textttOrig=\texttt
\def\texttt#1{\expandafter\textttOrig{\seqsplit{#1}}}
\renewcommand{\seqinsert}{\ifmmode
  \allowbreak
  \else\penalty6000\hspace{0pt plus 0.02em}\fi}

\newlength{\cslhangindent}
\newlength{\csllabelwidth}
\newenvironment{CSLReferences}[2] 
 {
  \setlength{\parindent}{0pt}
  \ifodd #1 \everypar{\setlength{\hangindent}{\cslhangindent}}\ignorespaces\fi
  \ifnum #2 > 0
  \setlength{\parskip}{#2\baselineskip}
  \fi
 }%
 {}
\usepackage{calc}

\usepackage[top=3.5cm, bottom=3cm, right=1.5cm, left=1.0cm,
            headheight=2.2cm, reversemp, includemp, marginparwidth=4.5cm]{geometry}

\titleformat{\section}
  {\normalfont\sffamily\Large\bfseries}
  {}{0pt}{}
\titleformat{\subsection}
  {\normalfont\sffamily\large\bfseries}
  {}{0pt}{}
\titleformat{\subsubsection}
  {\normalfont\sffamily\bfseries}
  {}{0pt}{}
\titleformat*{\paragraph}
  {\sffamily\normalsize}

\usepackage{fancyhdr}
\makeatletter
\let\ps@plain\ps@fancy
\fancyheadoffset[L]{4.5cm}
\fancyfootoffset[L]{4.5cm}

\definecolor{linky}{rgb}{0.0, 0.5, 1.0}

\newtcolorbox{repobox}
   {colback=red, colframe=red!75!black,
     boxrule=0.5pt, arc=2pt, left=6pt, right=6pt, top=3pt, bottom=3pt}

\newcommand{\ExternalLink}{%
   \tikz[x=1.2ex, y=1.2ex, baseline=-0.05ex]{%
       \begin{scope}[x=1ex, y=1ex]
           \clip (-0.1,-0.1)
               --++ (-0, 1.2)
               --++ (0.6, 0)
               --++ (0, -0.6)
               --++ (0.6, 0)
               --++ (0, -1);
           \path[draw,
               line width = 0.5,
               rounded corners=0.5]
               (0,0) rectangle (1,1);
       \end{scope}
       \path[draw, line width = 0.5] (0.5, 0.5)
           -- (1, 1);
       \path[draw, line width = 0.5] (0.6, 1)
           -- (1, 1) -- (1, 0.6);
       }
   }

\patchcmd{\@maketitle}{center}{flushleft}{}{}
\patchcmd{\@maketitle}{center}{flushleft}{}{}
\patchcmd{\@maketitle}{\LARGE}{\LARGE\sffamily}{}{}
\def\maketitle{{%
  
  \AB@maketitle}}
\makeatletter
\renewcommand\AB@affilsepx{ \protect\Affilfont}
\renewcommand\AB@affilnote[1]{{\bfseries #1}\hspace{3pt}}
\renewcommand{\affil}[2][]%
   {\newaffiltrue\let\AB@blk@and\AB@pand
      \if\relax#1\relax\def\AB@note{\AB@thenote}\else\def\AB@note{#1}%
        \setcounter{Maxaffil}{0}\fi
        \begingroup
        \let\href=\href@Orig
        \let\texttt=\textttOrig
        \let\protect\@unexpandable@protect
        \def\thanks{\protect\thanks}\def\footnote{\protect\footnote}%
        \@temptokena=\expandafter{\AB@authors}%
        {\def\\{\protect\\\protect\Affilfont}\xdef\AB@temp{#2}}%
         \xdef\AB@authors{\the\@temptokena\AB@las\AB@au@str
         \protect\\[\affilsep]\protect\Affilfont\AB@temp}%
         \gdef\AB@las{}\gdef\AB@au@str{}%
        {\def\\{, \ignorespaces}\xdef\AB@temp{#2}}%
        \@temptokena=\expandafter{\AB@affillist}%
        \xdef\AB@affillist{\the\@temptokena \AB@affilsep
          \AB@affilnote{\AB@note}\protect\Affilfont\AB@temp}%
      \endgroup
       \let\AB@affilsep\AB@affilsepx
}
\makeatother

\renewcommand\Affilfont{\sffamily\small\mdseries}
\ifnum 0\ifxetex 1\fi\ifluatex 1\fi=0 
  \usepackage[T1]{fontenc}
  \usepackage[utf8]{inputenc}

\else 
  \ifxetex
    \usepackage{mathspec}
    \usepackage{fontspec}

  \else
    \usepackage{fontspec}
  \fi
  \defaultfontfeatures{Ligatures=TeX,Scale=MatchLowercase}

\fi
\IfFileExists{upquote.sty}{\usepackage{upquote}}{}
\IfFileExists{microtype.sty}{%
\usepackage{microtype}
\UseMicrotypeSet[protrusion]{basicmath} 
}{}

\PassOptionsToPackage{usenames,dvipsnames}{color} 
\ifLuaTeX
\usepackage[bidi=basic]{babel}
\else
\usepackage[bidi=default]{babel}
\fi
\babelprovide[main,import]{american}

\def\languageshorthands#1{}

\let\addcontentslineOrig=\addcontentsline
\def\addcontentsline#1#2#3{\bgroup
  \let\texttt=\textttOrig\addcontentslineOrig{#1}{#2}{#3}\egroup}
\let\markbothOrig\markboth
\def\markboth#1#2{\bgroup
  \let\texttt=\textttOrig\markbothOrig{#1}{#2}\egroup}
\let\markrightOrig\markright
\def\markright#1{\bgroup
  \let\texttt=\textttOrig\markrightOrig{#1}\egroup}

\usepackage{graphicx,grffile}
\makeatletter
\def\maxwidth{\ifdim\Gin@nat@width>\linewidth\linewidth\else\Gin@nat@width\fi}
\def\maxheight{\ifdim\Gin@nat@height>\textheight\textheight\else\Gin@nat@height\fi}
\makeatother
\setkeys{Gin}{width=\maxwidth,height=\maxheight,keepaspectratio}
\IfFileExists{parskip.sty}{%
\usepackage{parskip}
}{
\setlength{\parindent}{0pt}
\setlength{\parskip}{6pt plus 2pt minus 1pt}
}
\providecommand{\tightlist}{%
  \setlength{\itemsep}{0pt}\setlength{\parskip}{0pt}}
\ifx\paragraph\undefined\else
\let\oldparagraph\paragraph
\renewcommand{\paragraph}[1]{\oldparagraph{#1}\mbox{}}
\fi
\ifx\subparagraph\undefined\else
\let\oldsubparagraph\subparagraph
\renewcommand{\subparagraph}[1]{\oldsubparagraph{#1}\mbox{}}
\fi

\title{\texttt{Eureka!}: An End-to-End Pipeline for JWST Time-Series
Observations}

\author[1%
]{Taylor J. Bell%
  \,\orcidlink{0000-0003-4177-2149}\,%
}
\author[2%
]{Eva-Maria Ahrer%
  \,\orcidlink{0000-0003-0973-8426}\,%
}
\author[3%
]{Jonathan Brande%
  \,\orcidlink{0000-0002-2072-6541}\,%
}
\author[4%
]{Aarynn L. Carter%
  \,\orcidlink{0000-0001-5365-4815}\,%
}
\author[5%
]{Adina D. Feinstein%
  \,\orcidlink{0000-0002-9464-8101}\,%
}
\author[6%
]{Giannina Guzman Caloca%
  \,\orcidlink{0000-0001-6340-8220}\,%
}
\author[7,8%
]{Megan Mansfield%
  \,\orcidlink{0000-0003-4241-7413}\,%
}
\author[9%
]{Sebastian Zieba%
  \,\orcidlink{0000-0003-0562-6750}\,%
}
\author[10%
]{Caroline Piaulet%
  \,\orcidlink{0000-0002-2875-917X}\,%
}
\author[10%
]{Björn Benneke%
  \,\orcidlink{0000-0001-5578-1498}\,%
}
\author[11%
]{Joseph Filippazzo%
  \,\orcidlink{0000-0002-0201-8306}\,%
}
\author[12%
]{Erin M. May%
  \,\orcidlink{0000-0002-2739-1465}\,%
}
\author[10%
]{Pierre-Alexis Roy%
  \,\orcidlink{0000-0001-6809-3520}\,%
}
\author[9%
]{Laura Kreidberg%
  \,\orcidlink{0000-0003-0514-1147}\,%
}
\author[12%
]{Kevin B. Stevenson%
  \,\orcidlink{0000-0002-7352-7941}\,%
}

\affil[1]{BAER Institute, NASA Ames Research Center, Moffet Field, CA
94035, USA}
\affil[2]{Department of Physics, University of Warwick, Gibbet Hill
Road, CV4 7AL Coventry, UK}
\affil[3]{Department of Physics and Astronomy, University of Kansas,
1082 Malott, 1251 Wescoe Hall Dr., Lawrence, KS 66045, USA}
\affil[4]{Department of Astronomy and Astrophysics, University of
California, Santa Cruz, 1156 High Street, Santa Cruz, CA 95064, USA}
\affil[5]{Department of Astronomy \& Astrophysics, University of
Chicago, 5640 S. Ellis Avenue, Chicago, IL 60637, USA}
\affil[6]{Department of Astronomy, University of Maryland, College Park,
MD USA}
\affil[7]{Steward Observatory, University of Arizona, Tucson, AZ 85719,
USA}
\affil[8]{NHFP Sagan Fellow}
\affil[9]{Max-Planck-Institut für Astronomie, Königstuhl 17, D-69117
Heidelberg, Germany}
\affil[10]{Department of Physics and Institute for Research on
Exoplanets, Université de Montréal, Montreal, QC, Canada}
\affil[11]{Space Telescope Science Institute, 3700 San Martin Drive,
Baltimore, MD 21218, USA}
\affil[12]{Johns Hopkins APL, 11100 Johns Hopkins Road, Laurel, MD
20723, USA}
\date{\vspace{-2.5ex}}

\begin{document}
\maketitle

\marginpar{

  \begin{flushleft}
  \sffamily\small

  {\bfseries DOI:} \href{https://doi.org/10.21105/joss.04503}{\color{linky}{10.21105/joss.04503}}

  \vspace{2mm}

  {\bfseries Software}
  \begin{itemize}
    \setlength\itemsep{0em}
    \item \href{https://github.com/openjournals/joss-reviews/issues/4503}{\color{linky}{Review}} \ExternalLink
    \item \href{https://github.com/kevin218/Eureka}{\color{linky}{Repository}} \ExternalLink
    \item \href{https://doi.org/10.5281/zenodo.7278300}{\color{linky}{Archive}} \ExternalLink
  \end{itemize}

  \vspace{2mm}

  \par\noindent\hrulefill\par

  \vspace{2mm}

  {\bfseries Editor:}\,\href{https://dfm.io/}{Dan\,Foreman-Mackey}\ExternalLink\orcidlink{0000-0002-9328-5652}
   \\
  \vspace{1mm}
    {\bfseries Reviewers:}
  \begin{itemize}
  \setlength\itemsep{0em}
    \item \href{https://github.com/catrionamurray}{@catrionamurray}
    \item \href{https://github.com/christinahedges}{@christinahedges}
    \item \href{https://github.com/dfm}{@dfm}
    \end{itemize}
    \vspace{2mm}

  {\bfseries Submitted:} 03 June 2022\\
  {\bfseries Published:} 03 November 2022

  \vspace{2mm}
  {\bfseries License}\\
  Authors of papers retain copyright and release the work under a Creative Commons Attribution 4.0 International License (\href{https://creativecommons.org/licenses/by/4.0/}{\color{linky}{CC BY 4.0}}).

  \end{flushleft}
}

\hypertarget{summary}{%
\section{Summary}\label{summary}}

\texttt{Eureka!} is a data reduction and analysis pipeline for exoplanet
time-series observations, with a particular focus on James Webb Space
Telescope (JWST, \protect\hyperlink{ref-JWST:2006}{Gardner et al.,
2006}) data. JWST was launched on December 25, 2021 and over the next
1-2 decades will pursue four main science themes: Early Universe,
Galaxies Over Time, Star Lifecycle, and Other Worlds. Our focus is on
providing the astronomy community with an open source tool for the
reduction and analysis of time-series observations of exoplanets in
pursuit of the fourth of these themes, Other Worlds. The goal of
\texttt{Eureka!} is to provide an end-to-end pipeline that starts with
raw, uncalibrated FITS files and ultimately yields precise exoplanet
transmission and/or emission spectra. The pipeline has a modular
structure with six stages, and each stage uses a “Eureka! Control File”
(ECF; these files use the .ecf file extension) to allow for easy control
of the pipeline’s behavior. Stage 5 also uses a “Eureka! Parameter File”
(EPF; these files use the .epf file extension) to control the fitted
parameters. We provide template ECFs for the MIRI
(\protect\hyperlink{ref-MIRI:2015}{Rieke et al., 2015}), NIRCam
(\protect\hyperlink{ref-NIRCam:2004}{Horner \& Rieke, 2004}), NIRISS
(\protect\hyperlink{ref-NIRISS:2017}{Maszkiewicz, 2017}), and NIRSpec
(\protect\hyperlink{ref-NIRSpec:2007}{Bagnasco et al., 2007})
instruments on JWST and the WFC3 instrument
(\protect\hyperlink{ref-WFC3:2008}{Kimble et al., 2008}) on the Hubble
Space Telescope (HST, \protect\hyperlink{ref-HST:1986}{Bahcall, 1986}).
These templates give users a good starting point for their analyses, but
\texttt{Eureka!} is not intended to be used as a black box tool, and
users should expect to fine-tune some settings for each observation in
order to achieve optimal results. At each stage, the pipeline creates
intermediate figures and outputs that allow users to compare
\texttt{Eureka!}’s performance using different parameter settings or to
compare \texttt{Eureka!} with an independent pipeline. The ECF used to
run each stage is also copied into the output folder from each stage to
enhance reproducibility. Finally, while \texttt{Eureka!} has been
optimized for exoplanet observations (especially the latter stages of
the code), much of the core functionality could also be repurposed for
JWST time-series observations in other research domains thanks to
\texttt{Eureka!}’s modularity.

\hypertarget{outline-of-eurekas-stages}{%
\section{\texorpdfstring{Outline of \texttt{Eureka!}’s
Stages}{Outline of Eureka!’s Stages}}\label{outline-of-eurekas-stages}}

\begin{figure}
\centering
\includegraphics[width=1\textwidth,height=\textheight]{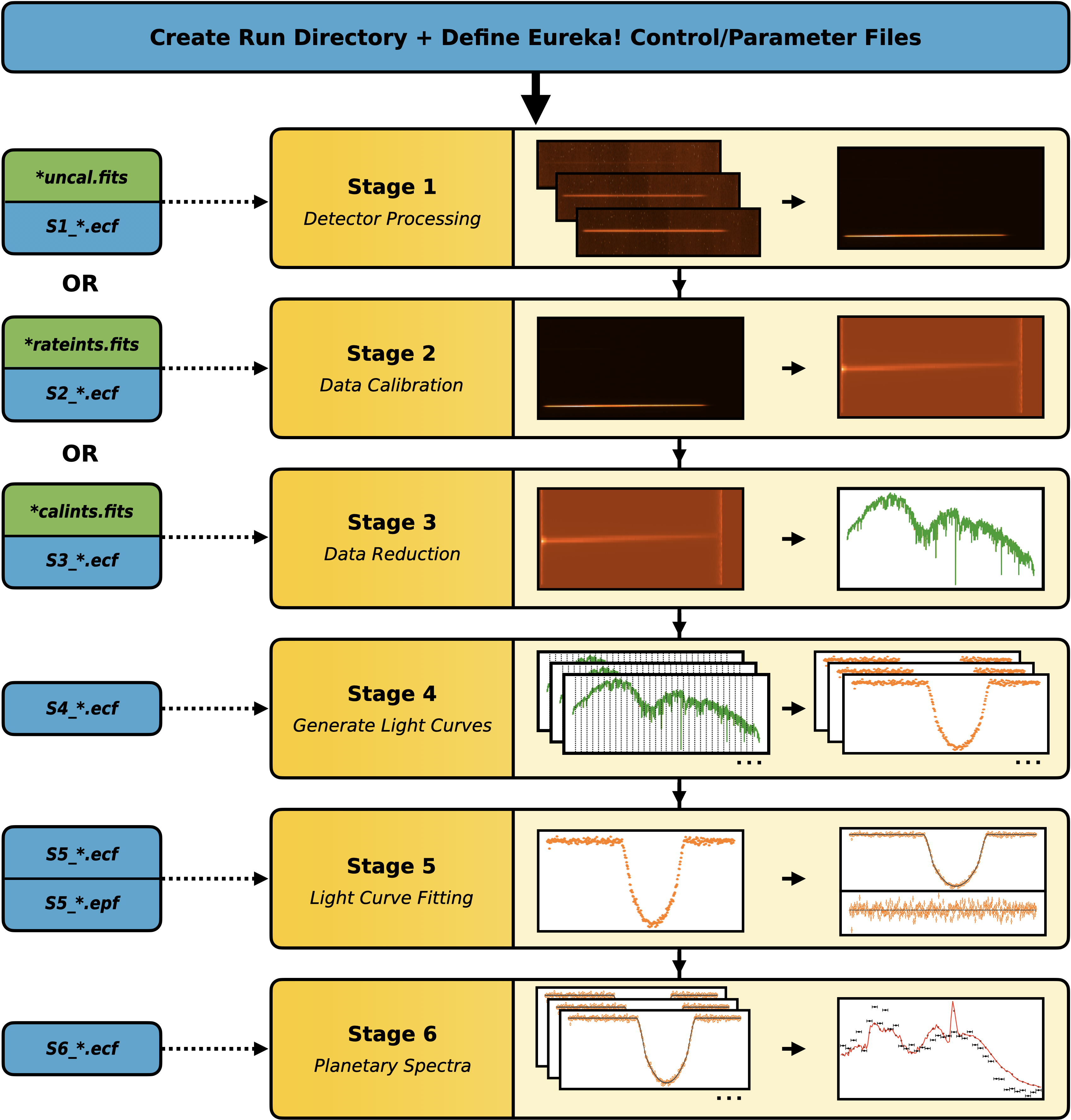}
\caption{An overview flowchart showing the processing done at each stage
in \texttt{Eureka!}. The outputs of each stage are used as the inputs to
the subsequent stage along with the relevant settings file(s).
\label{fig:overview}}
\end{figure}

\texttt{Eureka!} is broken down into six stages, which are as follows
(also summarized in \autoref{fig:overview}):

\begin{itemize}
\tightlist
\item
  Stage 1: An optional step that calibrates raw data (converts ramps to
  slopes for JWST observations). This step can be skipped within
  \texttt{Eureka!} if you would rather use the Stage 1 outputs from the
  \texttt{jwst} pipeline (\protect\hyperlink{ref-jwst:2022}{Bushouse et
  al., 2022}).
\item
  Stage 2: An optional step that further calibrates Stage 1 data
  (performs flat-fielding, unit conversion, etc. for JWST observations).
  This step can be skipped within \texttt{Eureka!} if you would rather
  use the Stage 2 outputs from the \texttt{jwst} pipeline.
\item
  Stage 3: Using Stage 2 outputs, performs background subtraction and
  optimal spectral extraction. For spectroscopic observations, this
  stage generates a time series of 1D spectra. For photometric
  observations, this stage generates a single light curve of flux versus
  time.
\item
  Stage 4: Using Stage 3 outputs, generates spectroscopic light curves
  by binning the time series of 1D spectra along the wavelength axis.
  Optionally removes drift/jitter along the dispersion direction and/or
  sigma clips outliers.
\item
  Stage 5: Fits the light curves with noise and astrophysical models
  using different optimization or sampling algorithms.
\item
  Stage 6: Displays the planet spectrum in figure and table form using
  results from the Stage 5 fits.
\end{itemize}

\hypertarget{differences-from-the-jwst-pipeline}{%
\section{\texorpdfstring{Differences From the \texttt{jwst}
Pipeline}{Differences From the jwst Pipeline}}\label{differences-from-the-jwst-pipeline}}

Eureka’s Stage 1 offers a few alternative, experimental ramp fitting
methods compared to the \texttt{jwst} pipeline, but mostly acts as a
wrapper to allow you to call the \texttt{jwst} pipeline in the same
format as \texttt{Eureka!}. Similarly, \texttt{Eureka!}’s Stage 2 acts
solely as a wrapper for the \texttt{jwst} pipeline. Meanwhile,
\texttt{Eureka!}’s Stages 3 through 6 completely depart from the
\texttt{jwst} pipeline and offer specialized background subtraction,
source extraction, wavelength binning, sigma clipping, fitting, and
plotting routines with heritage from past space-based exoplanet science.

\hypertarget{statement-of-need}{%
\section{Statement of Need}\label{statement-of-need}}

The calibration, reduction, and fitting of exoplanet time-series
observations is a challenging problem with many tunable parameters
across many stages, many of which will significantly impact the final
results. Typically, the default calibration pipeline from astronomical
observatories is insufficiently tailored for exoplanet time-series
observations as the pipeline is more optimized for other science use
cases. As such, it is common practice to develop a custom data analysis
pipeline that starts from the original, uncalibrated images.
Historically, data analysis pipelines have often been proprietary, so
each new user of an instrument or telescope has had to develop their own
pipeline. Also, clearly specifying the analysis procedure can be
challenging, especially with proprietary code, which erodes
reproducibility. \texttt{Eureka!} seeks to be a next-generation data
analysis pipeline for next-generation observations from JWST with
open-source and well-documented code for easier adoption; modular code
for easier customization while maintaining a consistent framework; and
easy-to-use but powerful inputs and outputs for increased automation,
increased reproducibility, and more thorough intercomparisons. By also
allowing for analyses of HST observations within the same framework,
users will be able to combine new and old observations to develop a more
complete understanding of individual targets or even entire populations.

\hypertarget{documentation}{%
\section{Documentation}\label{documentation}}

Documentation for \texttt{Eureka!} is available at
\url{https://eurekadocs.readthedocs.io/en/latest/}.

\hypertarget{similar-tools}{%
\section{Similar Tools}\label{similar-tools}}

We will now discuss the broader data reduction and fitting ecosystem in
which \texttt{Eureka!} lives. Several similar open-source tools are
discussed below to provide additional context, but this is not meant to
be a comprehensive list.

As mentioned above, \texttt{Eureka!} makes use of the first two stages
of \href{https://github.com/spacetelescope/jwst}{\texttt{jwst}}
(\protect\hyperlink{ref-jwst:2022}{Bushouse et al., 2022}) while
offering significantly different extraction routines and novel spectral
binning and fitting routines beyond what is contained in \texttt{jwst}.
\texttt{Eureka!} bears similarities to the
\href{https://github.com/kevin218/POET}{\texttt{POET}}
(\protect\hyperlink{ref-Cubillos:2013}{Cubillos et al., 2013};
\protect\hyperlink{ref-Stevenson:2012}{Stevenson et al., 2012}) and
\href{https://github.com/kevin218/WFC3}{\texttt{WFC3}}
(\protect\hyperlink{ref-Stevenson:2014a}{Stevenson et al., 2014})
pipelines, developed for Spitzer/IRAC and HST/WFC3 observations
respectively; in fact, much of the code from those pipelines has been
incorporated into \texttt{Eureka!}. \texttt{Eureka!} is near feature
parity with \texttt{WFC3}, but the Spitzer specific parts of the
\texttt{POET} pipeline have not been encorporated into \texttt{Eureka!}.
The \href{https://github.com/lisadang27/SPCA}{\texttt{SPCA}}
(\protect\hyperlink{ref-Bell:2021}{Bell et al., 2021};
\protect\hyperlink{ref-Dang:2018}{Dang et al., 2018}) pipeline developed
for the reduction and fitting of Spitzer/IRAC observations also bears
some similarity to this pipeline, and some snippets of that pipeline
have also been encorporated into \texttt{Eureka!}. The
\href{https://github.com/eas342/tshirt}{\texttt{tshirt}}
(\protect\hyperlink{ref-tshirt:2022}{Schlawin \& Glidic, 2022}) package
also offers spectral and photometric extraction routines that work for
HST and JWST data.
\href{https://github.com/sebastian-zieba/PACMAN}{\texttt{PACMAN}}
(\protect\hyperlink{ref-Kreidberg:2014}{Kreidberg et al., 2014};
\protect\hyperlink{ref-pacman:2022}{Zieba \& Kreidberg, 2022}) is
another open-source end-to-end pipeline developed for HST/WFC3
observations. The
\href{https://github.com/exoplanet-dev/exoplanet}{\texttt{exoplanet}}
(\protect\hyperlink{ref-exoplanet:2021}{Foreman-Mackey et al., 2021})
and \href{https://github.com/nespinoza/juliet}{\texttt{juliet}}
(\protect\hyperlink{ref-juliet:2019}{Espinoza et al., 2019}) packages
offer some similar capabilities as the observation fitting parts of
\texttt{Eureka!}.

\hypertarget{acknowledgements}{%
\section{Acknowledgements}\label{acknowledgements}}

\texttt{Eureka!} allows for some variations upon the STScI’s
\href{https://github.com/spacetelescope/jwst}{\texttt{jwst}} pipeline
(\protect\hyperlink{ref-jwst:2022}{Bushouse et al., 2022}) for Stages 1
and 2, but presently these stages mostly act as wrappers around the
\texttt{jwst} pipeline. This allows \texttt{Eureka!} to run the
\texttt{jwst} pipeline in the same manner as \texttt{Eureka!}’s latter
stages. \texttt{Eureka!} then uses its own custom code for additional
calibration steps, spectral or photometric extraction, and light curve
fitting. Several parts of the spectroscopy-focused code in Stages 3 and
4 of \texttt{Eureka!} were inspired by, or were initially written for,
the \href{https://github.com/kevin218/WFC3}{\texttt{WFC3}}
(\protect\hyperlink{ref-Stevenson:2014a}{Stevenson et al., 2014})
pipeline. Other parts of the spectroscopy code and several parts of the
photometry focused code in Stage 3 were inspired by, or were initially
written for, the \href{https://github.com/kevin218/POET}{\texttt{POET}}
pipeline (\protect\hyperlink{ref-Cubillos:2013}{Cubillos et al., 2013};
\protect\hyperlink{ref-Stevenson:2012}{Stevenson et al., 2012}). Some of
the Stage 5 code comes from Kreidberg et al.
(\protect\hyperlink{ref-Kreidberg:2014}{2014}) and
\href{https://github.com/sebastian-zieba/PACMAN}{\texttt{PACMAN}}
(\protect\hyperlink{ref-pacman:2022}{Zieba \& Kreidberg, 2022}). Small
pieces of the \href{https://github.com/lisadang27/SPCA}{\texttt{SPCA}}
(\protect\hyperlink{ref-Bell:2021}{Bell et al., 2021};
\protect\hyperlink{ref-Dang:2018}{Dang et al., 2018}) and
\href{https://github.com/taylorbell57/Bell_EBM}{\texttt{Bell\_EBM}}
(\protect\hyperlink{ref-Bell:2018}{Bell \& Cowan, 2018}) repositories
have also been reused.

ALC is supported by a grant from STScI (\emph{JWST}-ERS-01386) under
NASA contract NAS5-03127. ADF acknowledges support by the National
Science Foundation Graduate Research Fellowship Program under Grant
No.~(DGE-1746045). CP acknowledges financial support by the Fonds de
Recherche Québécois—Nature et Technologie (FRQNT; Québec), the
Technologies for Exo-Planetary Science (TEPS) Trainee Program and the
Natural Sciences and Engineering Research Council (NSERC) Vanier
Scholarship. JB acknowledges support from the NASA Interdisciplinary
Consortia for Astrobiology Research (ICAR). KBS is supported by
\emph{JWST}-ERS-01366. MM acknowledges support through the NASA Hubble
Fellowship grant HST-HF2-51485.001-A awarded by STScI, which is operated
by the Association of Universities for Research in Astronomy, Inc., for
NASA, under contract NAS5-26555. We also thank Ivelina Momcheva for
useful discussions. Support for this work was provided in part by NASA
through a grant from the Space Telescope Science Institute, which is
operated by the Association of Universities for Research in Astronomy,
Inc., under NASA contract NAS 5-03127. In addition, we would like to
thank the Transiting Exoplanet Community Early Release Science program
for organizing meetings that contributed to the writing of
\texttt{Eureka!}.

\hypertarget{references}{%
\section*{References}\label{references}}
\addcontentsline{toc}{section}{References}

\hypertarget{refs}{}
\begin{CSLReferences}{1}{0}
\leavevmode\vadjust pre{\hypertarget{ref-NIRSpec:2007}{}}%
Bagnasco, G., Kolm, M., Ferruit, P., Honnen, K., Koehler, J., Lemke, R.,
Maschmann, M., Melf, M., Noyer, G., Rumler, P., Salvignol, J.-C.,
Strada, P., \& Te Plate, M. (2007). {Overview of the near-infrared
spectrograph (NIRSpec) instrument on-board the James Webb Space
Telescope (JWST)}. In J. B. Heaney \& L. G. Burriesci (Eds.),
\emph{Cryogenic optical systems and instruments XII} (Vol. 6692, p.
66920M). \url{https://doi.org/10.1117/12.735602}

\leavevmode\vadjust pre{\hypertarget{ref-HST:1986}{}}%
Bahcall, N. A. (1986). {The Hubble Space Telescope.} \emph{Annals of the
New York Academy of Sciences}, \emph{470}, 331–337.
\url{https://doi.org/10.1111/j.1749-6632.1986.tb47983.x}

\leavevmode\vadjust pre{\hypertarget{ref-Bell:2018}{}}%
Bell, T. J., \& Cowan, N. B. (2018). {Increased Heat Transport in
Ultra-hot Jupiter Atmospheres through H\(_{2}\) Dissociation and
Recombination}. \emph{ApJL}, \emph{857}(2), L20.
\url{https://doi.org/10.3847/2041-8213/aabcc8}

\leavevmode\vadjust pre{\hypertarget{ref-Bell:2021}{}}%
Bell, T. J., Dang, L., Cowan, N. B., Bean, J., Désert, J.-M., Fortney,
J. J., Keating, D., Kempton, E., Kreidberg, L., Line, M. R., Mansfield,
M., Parmentier, V., Stevenson, K. B., Swain, M., \& Zellem, R. T.
(2021). {A comprehensive reanalysis of Spitzer’s 4.5 {\(\mu\)}m phase
curves, and the phase variations of the ultra-hot Jupiters MASCARA-1b
and KELT-16b}. \emph{MNRAS}, \emph{504}(3), 3316–3337.
\url{https://doi.org/10.1093/mnras/stab1027}

\leavevmode\vadjust pre{\hypertarget{ref-jwst:2022}{}}%
Bushouse, H., Eisenhamer, J., Dencheva, N., Davies, J., Greenfield, P.,
Morrison, J., Hodge, P., Simon, B., Grumm, D., Droettboom, M., Slavich,
E., Sosey, M., Pauly, T., Miller, T., Jedrzejewski, R., Hack, W., Davis,
D., Crawford, S., Law, D., … Jamieson, W. (2022). \emph{{JWST
Calibration Pipeline 1.6.2}} (Version 1.6.2). Zenodo.
\url{https://doi.org/10.5281/zenodo.6984366}

\leavevmode\vadjust pre{\hypertarget{ref-Cubillos:2013}{}}%
Cubillos, P., Harrington, J., Madhusudhan, N., Stevenson, K. B., Hardy,
R. A., Blecic, J., Anderson, D. R., Hardin, M., \& Campo, C. J. (2013).
{WASP-8b: Characterization of a Cool and Eccentric Exoplanet with
Spitzer}. \emph{ApJ}, \emph{768}(1), 42.
\url{https://doi.org/10.1088/0004-637X/768/1/42}

\leavevmode\vadjust pre{\hypertarget{ref-Dang:2018}{}}%
Dang, L., Cowan, N. B., Schwartz, J. C., Rauscher, E., Zhang, M.,
Knutson, H. A., Line, M., Dobbs-Dixon, I., Deming, D., Sundararajan, S.,
Fortney, J. J., \& Zhao, M. (2018). {Detection of a westward hotspot
offset in the atmosphere of hot gas giant CoRoT-2b}. \emph{Nature
Astronomy}, \emph{2}, 220–227.
\url{https://doi.org/10.1038/s41550-017-0351-6}

\leavevmode\vadjust pre{\hypertarget{ref-juliet:2019}{}}%
Espinoza, N., Kossakowski, D., \& Brahm, R. (2019). {juliet: a versatile
modelling tool for transiting and non-transiting exoplanetary systems}.
\emph{MNRAS}, \emph{490}(2), 2262–2283.
\url{https://doi.org/10.1093/mnras/stz2688}

\leavevmode\vadjust pre{\hypertarget{ref-exoplanet:2021}{}}%
Foreman-Mackey, D., Luger, R., Agol, E., Barclay, T., Bouma, L. G.,
Brandt, T. D., Czekala, I., David, T. J., Dong, J., Gilbert, E. A.,
Gordon, T. A., Hedges, C., Hey, D. R., Morris, B. M., Price-Whelan, A.
M., \& Savel, A. B. (2021). \emph{{exoplanet: Gradient-based
probabilistic inference for exoplanet data \& other astronomical time
series}} (Version 0.5.1). Zenodo; Zenodo.
\url{https://doi.org/10.5281/zenodo.1998447}

\leavevmode\vadjust pre{\hypertarget{ref-JWST:2006}{}}%
Gardner, J. P., Mather, J. C., Clampin, M., Doyon, R., Greenhouse, M.
A., Hammel, H. B., Hutchings, J. B., Jakobsen, P., Lilly, S. J., Long,
K. S., Lunine, J. I., McCaughrean, M. J., Mountain, M., Nella, J.,
Rieke, G. H., Rieke, M. J., Rix, H.-W., Smith, E. P., Sonneborn, G., …
Wright, G. S. (2006). {The James Webb Space Telescope}. \emph{123}(4),
485–606. \url{https://doi.org/10.1007/s11214-006-8315-7}

\leavevmode\vadjust pre{\hypertarget{ref-NIRCam:2004}{}}%
Horner, S. D., \& Rieke, M. J. (2004). {The near-infrared camera
(NIRCam) for the James Webb Space Telescope (JWST)}. In J. C. Mather
(Ed.), \emph{Optical, infrared, and millimeter space telescopes} (Vol.
5487, pp. 628–634). \url{https://doi.org/10.1117/12.552281}

\leavevmode\vadjust pre{\hypertarget{ref-WFC3:2008}{}}%
Kimble, R. A., MacKenty, J. W., O’Connell, R. W., \& Townsend, J. A.
(2008). {Wide Field Camera 3: a powerful new imager for the Hubble Space
Telescope}. In Jr. Oschmann Jacobus M., M. W. M. de Graauw, \& H. A.
MacEwen (Eds.), \emph{Space telescopes and instrumentation 2008:
Optical, infrared, and millimeter} (Vol. 7010, p. 70101E).
\url{https://doi.org/10.1117/12.789581}

\leavevmode\vadjust pre{\hypertarget{ref-Kreidberg:2014}{}}%
Kreidberg, L., Bean, J. L., Désert, J.-M., Benneke, B., Deming, D.,
Stevenson, K. B., Seager, S., Berta-Thompson, Z., Seifahrt, A., \&
Homeier, D. (2014). {Clouds in the atmosphere of the super-Earth
exoplanet GJ 1214b}. \emph{Nature}, \emph{505}(7481), 69–72.
\url{https://doi.org/10.1038/nature12888}

\leavevmode\vadjust pre{\hypertarget{ref-NIRISS:2017}{}}%
Maszkiewicz, M. (2017). {Near- infrared imager and slitless spectrograph
(NIRISS): a new instrument on James Webb Space Telescope (JWST)}.
\emph{Society of Photo-Optical Instrumentation Engineers (SPIE)
Conference Series}, \emph{10564}, 105642Q.
\url{https://doi.org/10.1117/12.2309161}

\leavevmode\vadjust pre{\hypertarget{ref-MIRI:2015}{}}%
Rieke, G. H., Wright, G. S., Böker, T., Bouwman, J., Colina, L., Glasse,
A., Gordon, K. D., Greene, T. P., Güdel, M., Henning, Th., Justtanont,
K., Lagage, P.-O., Meixner, M. E., Nørgaard-Nielsen, H.-U., Ray, T. P.,
Ressler, M. E., van Dishoeck, E. F., \& Waelkens, C. (2015). {The
Mid-Infrared Instrument for the James Webb Space Telescope, I:
Introduction}. \emph{127}(953), 584.
\url{https://doi.org/10.1086/682252}

\leavevmode\vadjust pre{\hypertarget{ref-tshirt:2022}{}}%
Schlawin, E., \& Glidic, K. (2022). {tshirt}. In \emph{GitHub
repository}. GitHub. \url{https://github.com/eas342/tshirt}

\leavevmode\vadjust pre{\hypertarget{ref-Stevenson:2014a}{}}%
Stevenson, K. B., Bean, J. L., Seifahrt, A., Désert, J.-M., Madhusudhan,
N., Bergmann, M., Kreidberg, L., \& Homeier, D. (2014). {Transmission
Spectroscopy of the Hot Jupiter WASP-12b from 0.7 to 5 {\(\mu\)}m}.
\emph{AJ}, \emph{147}(6), 161.
\url{https://doi.org/10.1088/0004-6256/147/6/161}

\leavevmode\vadjust pre{\hypertarget{ref-Stevenson:2012}{}}%
Stevenson, K. B., Harrington, J., Fortney, J. J., Loredo, T. J., Hardy,
R. A., Nymeyer, S., Bowman, W. C., Cubillos, P., Bowman, M. O., \&
Hardin, M. (2012). {Transit and Eclipse Analyses of the Exoplanet HD
149026b Using BLISS Mapping}. \emph{ApJ}, \emph{754}(2), 136.
\url{https://doi.org/10.1088/0004-637X/754/2/136}

\leavevmode\vadjust pre{\hypertarget{ref-pacman:2022}{}}%
Zieba, S., \& Kreidberg, L. (2022). {PACMAN}. In \emph{GitHub
repository}. GitHub. \url{https://github.com/sebastian-zieba/PACMAN}

\end{CSLReferences}

\end{document}